\documentclass[draft]{agujournal2019}
\usepackage{url}
\usepackage{lineno}
\usepackage[inline]{trackchanges} 
\usepackage{soul}
\usepackage{float}
\usepackage{amsmath} 

\draftfalse
\journalname{Geophysical Research Letters}

\begin{document}
\title{Skillful Global Ocean Emulation and the Role of Correlation-Aware Loss}

\authors{N. Agarwal\affil{1,2}, T. A. Smith\affil{2}, S. Frolov\affil{2}, L. C. Slivinski\affil{2}}

\affiliation{1}{CIRES, University of Colorado, Boulder, CO, USA}
\affiliation{2}{NOAA Physical Sciences Laboratory, Boulder, CO, USA}

\correspondingauthor{Niraj Agarwal}{niraj.agarwal@colorado.edu}

\begin{keypoints}
\item Emulating global oceans on 24-hour time steps provides skillful medium-range forecasts without autoregressive training.
\item Training emulator using Mahalanobis distance as loss improves the forecast skill compared to training using mean squared error.
\item Considering target correlations explicitly regularizes the problem and improves spatial correlations amongst variables.
\end{keypoints}

\begin{abstract}
Machine learning emulators have shown extraordinary skill in forecasting atmospheric states, and their application to global ocean dynamics offers similar promise. 
Here, we adapt the GraphCast architecture into a dedicated ocean-only emulator, driven by prescribed atmospheric conditions, for medium-range predictions.
The emulator is trained on NOAA’s UFS-Replay dataset.
Using a $24$ hour time step, single initial condition, and without using autoregressive training, we produce an emulator that provides skillful forecasts for $10–15$ day lead times. 
We further demonstrate the use of Mahalanobis distance as loss that improves the forecast skill compared to the Mean Squared Error loss by explicitly accounting for the correlations between tendencies of the target variables.
Using spatial correlation analysis of the forecasted fields, we also show that the proposed correlation-aware loss acts as a statistical-dynamical regularizer for the slow, correlated dynamics of the global oceans, offering a better background forecast for downstream tasks like data assimilation.
\end{abstract}

\section*{Plain Language Summary}
Artificial intelligence (AI) has revolutionized numerical weather prediction; we use it here to forecast the oceans. 
We adapted a leading AI weather model, GraphCast, to simulate ocean conditions, such as, temperature, salinity, and currents up to $500$m depth, by training it on the high-quality UFS-Replay dataset of ocean states produced by NOAA. 
The emulator produced skillful forecasts on the intended $10$-days lead time for hundreds of initial conditions in the year 2022.
One of our key advancements is the use of a specialized training method that informed the emulator how different ocean properties are interrelated. 
This approach, which recognizes dominant physical and thermodynamic relationships in physical oceanography, improved the physical realism of the forecasts. 
Our work also uncovered an effective recipe for building skillful ocean emulators: choose the right time step for the architecture, use single initial condition, select distinct non-correlated features in the input space, and train by explicitly respecting the non-negligible inter-variable correlations in the target variables.
These findings pave the way for developing efficient emulators for the ocean-atmosphere coupled setup, where such interconnections in the forecasted variables are eminent and must be resolved to use them in operational state estimation tasks.

\section{Introduction}
The tremendous success of data-driven methods in learning the evolution of atmospheric states has brought a fundamental paradigm shift in weather forecasting, with several deterministic emulators now rivaling the skill of state-of-the-art global numerical weather prediction (NWP) models \cite{keisler2022forecasting,pathak2022fourcastnet,lam2023learning,bi2023accurate,lang2024aifs}. 
This success has naturally stimulated interest in developing similar capabilities for the chaotic dynamical system of Earth's oceans. 
Recent efforts have explored medium-range ocean emulation at $10–15$ day lead times on regional \cite{chattopadhyay2024oceannet,lupin2025simultaneous} and global scales \cite{el2025glonet,wang2024xihe,cui2025forecasting}, with some work also done on emulating lower-frequency variability using 5-day mean states for climate applications \cite{dheeshjith2025samudra}.
These ocean emulators, often trained on a coarsened version of $1^{\circ}/12$ GLORYS reanalysis \cite{lellouche2021copernicus}, have demonstrated an impressive capacity to resolve complex, non-linear, multiscale features such as mesoscale eddies and western boundary currents. 
Additionally, the several orders of magnitude gain in computational speedup offered by these emulators promises large ensemble forecasting and the potential to replace expensive numerical ocean models in coupled ocean-atmosphere forecasting for medium-range weather. 

This work is motivated by the broader objective of developing a fully coupled earth system emulator by including 3D oceans alongside a 3D atmosphere to achieve joint training within frameworks like GraphCast \cite{lam2023learning}.
However, a simple extension of atmospheric emulators to include 3D oceans is hindered by distinct physical challenges. 
Unlike the atmosphere, oceans operate on significantly different spatiotemporal scales with a much longer memory, i.e., slowly evolving dynamics. 
Therefore, the available training dataset for oceans contains fewer dynamical cycles of low-frequency variability than its atmospheric counterpart, and this may hinder robust training.  
Ocean emulation is additionally constrained by complexities due to continental boundaries, bathymetry, substantially different deeper ocean properties, and their communication with the upper ocean and the atmosphere. 
Understanding such issues and finding ways to improve the skill of oceanic emulators forms the core of this research. 
Here, we particularly focus on the assumption of statistical independence between forecasted variables, meaning all forecasted quantities are distinct from each other with zero cross-variable correlations.
This is clearly not the case, and the expectation that the network will learn this implicitly does not seem to hold given the poor performance of downstream tasks, such as data assimilation (DA) where resolving such inter-variable correlations is crucial for projecting information from observed to non-observed variables \cite{slivinski2025assimilating}.
This requirement would only get more important as the community moves towards building robust strongly-coupled data assimilation for integrated earth system analysis \cite{penny2017coupled}.

In this study, we attempt to improve the representation of inter-variable correlations using a correlation-aware loss function based on the Mahalanobis distance \cite{mahalanobis2018generalized}.
The Mahalanobis loss (hereafter, M-Loss) accounts for the anisotropic nature of forecast errors by utilizing a transformed set of principal axes, contrasting with the inherent isotropic error assumption in the standard Mean Squared Error (MSE).
To demonstrate the efficacy of this error metric in training, we first adapt GraphCast to produce a baseline MSE-based 3D ocean emulator that achieves skillful medium-range forecast and shows numerically stable roll out for several months without requiring multi-step rollout training.
We then retrain this architecture using M-Loss as the error metric for minimization. 
The results, although presented for oceanic emulators, also have utility for atmospheric and coupled ocean-atmosphere emulation where similar inter-variable correlations are also present and may benefit from their explicit accounting.

\section{Data and Methods}

\subsection{Dataset and Preprocessing}
\label{subsec:dataset_preprocess}
The ocean-only emulator presented in this study is trained using the reanalysis-like Unified Forecast System (UFS)-Replay dataset \cite{UFSReplay}. 
This dataset is generated by nudging high-resolution UFS model outputs to external reanalyses -- ERA5 \cite{hersbach2020era5} for the atmosphere and ORAS5 \cite{zuo2019ecmwf} for the oceans. 
The original UFS-Replay dataset is available from January $1994$ to October $2023$ at $0.25^{\circ}$ nominal spatial resolution, but here we subsampled it to $1^{\circ}$ after converting it to the Gaussian grid. 
The vertical subsampling is performed using layer thickness weighted averaging, a method conceptually similar to the pressure thickness weighted averaging utilized in ACE \cite{watt2023ace, watt2025ace2} climate emulator. 
More details on UFS-Replay and the vertical subsampling is discussed in supplementary text S1. 

\subsection{Emulator Design and Model Setup}
\label{subsec:problem_setup}
We represent the global ocean as a graph-based dynamical system by adapting the GraphCast architecture \cite{lam2023learning} to a 3D oceanic configuration.
Due to the distinct characteristics of oceanic flows and the dataset, several modifications were made to the GraphCast framework to produce a physically consistent ocean emulator.
The emulator, $f$, learns a mapping from the current ocean state ($\textbf{X}_t$) and prescribed future atmospheric boundary conditions ($\mathbf{F}_{t+1}$) to the deviations in the state at the next time step ($\Delta \mathbf{X}_{t+\Delta t}$), given as, 
\begin{eqnarray}
    \Delta \mathbf{X}_{t+\Delta t} & = & f(\mathbf{X}_t, \mathbf{F}_{t+\Delta t}), \hspace{5pt} \\
    \text{and} \hspace{8pt} \mathbf{X}_{t+\Delta t}  & = & \mathbf{X}_t + \Delta \mathbf{X}_{t+\Delta t} 
\end{eqnarray}
Here $\Delta t$ represents the emulator time step.
Note, that our configuration utilizes only one time step as the initial condition ($t$), rather than the two-step history used in the original GraphCast. 
Sensitivity tests using two time step inputs for emulators with $\Delta t = 6$ and $24$ hours showed significant grid imprinting -- honeycomb-like grid-locked structures -- in forecasts during autoregressive rollouts to $5$ days (see Figure S1). 
Results (not shown) suggest that this grid imprinting with two time steps is likely due to a rank-deficient or ill-conditioned input feature matrix, meaning the effective dimensionality of the input is much lower than the total number of features. 
Based on a recent study which provides a mechanistic understanding of GraphCast \cite{macmillan2025towards}, a likely explanation is that, in case of $2$ ICs and/or extensively correlated input features, the encoder possibly projects redundant features onto the latent space as a set of grid-locked noise, which compounds during autoregressive inference. 
Using a single $24$-hour time step effectively regularizes the input space and boosts the signal-to-noise ratio, ensuring a more robust projection into the latent manifold of the same dimension.
Whether these artifacts are architecture-dependent remains unclear and requires further investigation.

In terms of the problem formulation, predicting tendencies (or, deviations) rather than full states offers two distinct advantages. 
First, temporal tendencies in oceanic fields exhibit more near-Gaussian distributions than the full states, which are heavily skewed and demonstrate a long tail, e.g., SSH. 
Therefore, the gaussianity assumption used during normalization, by subtracting the mean and dividing by the standard deviation, aligns better with tendencies than the full state.
Second, the residual formulation explicitly adds a linear piece, using a skip connection, to the nonlinear component provided by the network, which has been shown to perform better in idealized oceanic emulation studies \cite{agarwal2021comparison}.

The emulator simulates a mix of 2D and 3D ocean states up to $500$m depth, driven by atmospheric boundary conditions. 
A comprehensive list of states and forcings used in the model setup is provided in supplementary Table S1.
The forcing variables are selected based on bulk flux formulations, but physical constants required for exact flux calculations are omitted, assuming the network implicitly internalizes these parameters.
We do not utilize diagnosed 2m temperature ($T_{2m}$) to avoid propagating the approximations inherent in its computation to the emulator. 
The interfaces of the seven vertical layers are placed at approximately $z$ $=$ $0$, $1$, $21$, $75$, $120$, $200$, $350$, $500$ meters below the sea level. 
A finer representation of the vertical direction using more vertical layers around the top $0-100$ meters (consistent with physical ocean models like MOM6) was also successfully trained/tested but is not reported here.
Finer vertical resolution in the mixed layer may similarly degrade the input matrix conditioning, but we found this to be less sensitive than the number of initial conditions.  

\subsection{Correlation-aware Loss}
\label{subsec:mahalanobis_loss}
To ensure physical consistency, we utilize a correlation-aware loss function based on the Mahalanobis distance. 
To our knowledge, this is the first application of such a loss in training a global Earth system emulator.
This loss relaxes the implicit assumption in MSE that forecast errors across different variables and vertical levels are independent and identically distributed. 
Geometrically, MSE defines an isotropic (spherical) error surface, penalizing deviations equally in all directions of the state space. 
However, ocean/atmospheric dynamics are characterized by strong multivariate couplings, such as the geostrophic relationship between sea surface height (SSH) and velocity (U, V), meaning that physically consistent errors lie on an ellipsoid rather than on sphere.
This anisotropic error structure is accounted for using the Mahalanobis distance which weights the loss function by the inverse of the inter-variable correlations. 
The loss function, $\mathcal{L}_M$, for a predicted increment, $\Delta \widehat{\mathbf{X}}$, relative to the ground-truth, $\Delta \mathbf{X}$, is given as,

\begin{equation}
\mathcal{L}_M = \sqrt{(\Delta \mathbf{X} - \Delta \widehat{\mathbf{X}}) \mathbf{\Sigma}^{-1} (\Delta \mathbf{X} - \Delta \widehat{\mathbf{X}})^\top}
\label{eq:maha_loss}
\end{equation}

\begin{figure}[ht]
    \centering
    \includegraphics[width=0.8\linewidth]{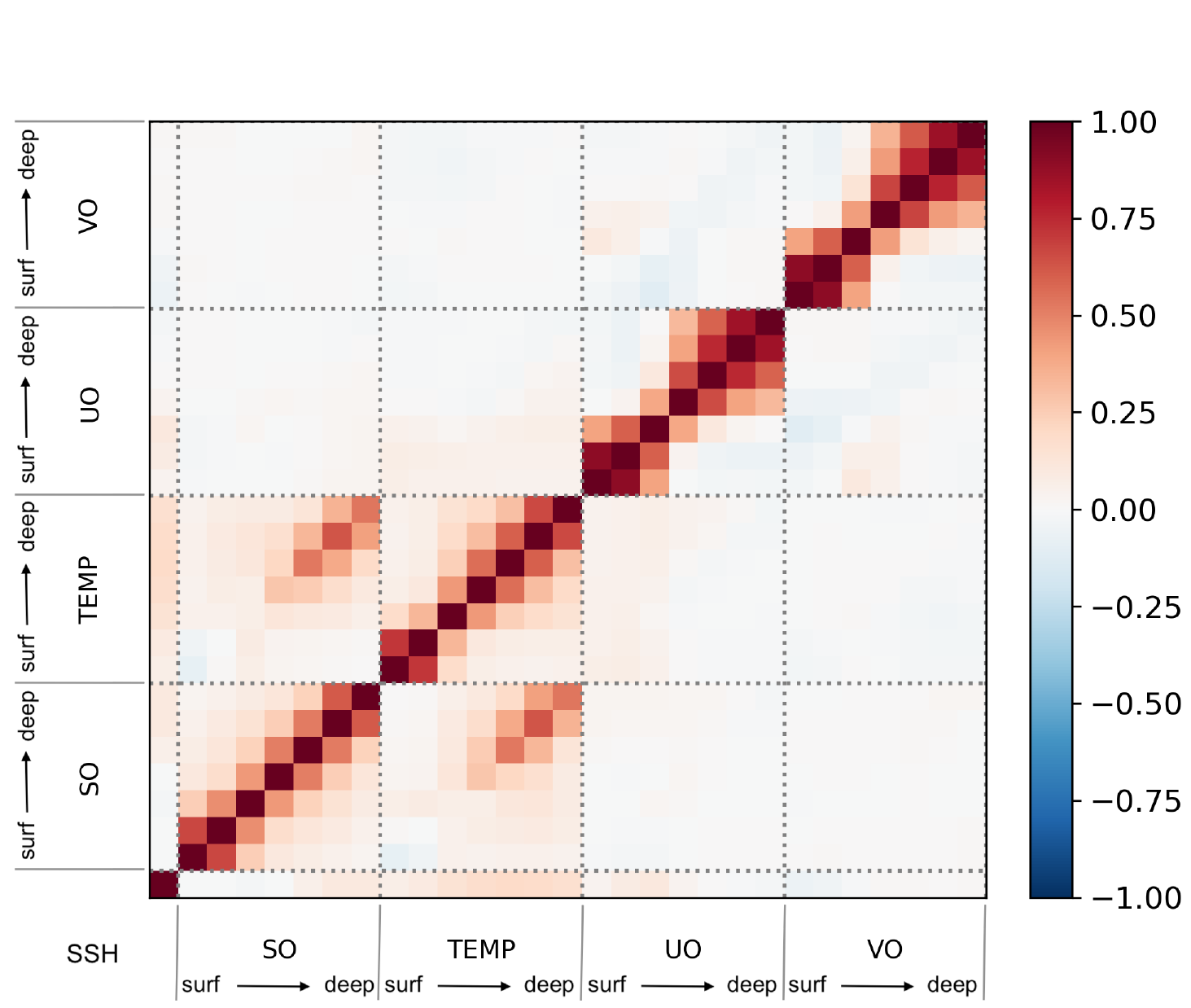}
    \caption{ Global inter-variable correlation matrix ($\mathbf\Sigma$) derived from the temporal tendencies of all prognostic variables. 
    For 3D fields (SO, TEMP, UO, VO), indices progress from surface to deep levels (left-to-right and bottom-to-top), while SSH is represented as a single entry in the bottom left corner. Diagonal elements are all one. The colorbar indicates Pearson correlation coefficients ranging between $[-1,1]$. }
    \label{fig:1}
\end{figure}

\noindent Here $\mathbf{\Sigma}$ is the correlations between all pairs $(i,j)$ of prognostic variables in their temporal tendencies over the training dataset ($1994-2019$), averaged over all grid points, i.e.,

\begin{eqnarray}
\mathbf{\Sigma}_{i,j} &=& \text{Corr}(\Delta \mathbf{X}_i, \Delta \mathbf{X}_j) 
\label{eq:corr_matrix1}
\\
& = & \left \langle \frac{\sum_{t=0}^{N-1} (X_i^{t+1} - X_i^t)(X_j^{t+1} - X_j^t)}
{\sqrt{\sum_{t=0}^{N-1} (X_i^{t+1} - X_i^t)^2}.\sqrt{\sum_{t=0}^{N-1} (X_j^{t+1} - X_j^t)^2}} \right \rangle_{(\text{lat},\text{lon})}
\label{eq:corr_matrix2}
\end{eqnarray}

\noindent where $\langle.\rangle$ denotes spatial averaging. 
Omitting the spatial averaging would produce separate $\mathbf{\Sigma}$ for each geographical location.
The globally-averaged $\mathbf \Sigma$ matrix computed for the problem setup described in Section \ref{subsec:problem_setup} is illustrated in Figure \ref{fig:1}. 
Note, that this is the simplest possible approximation of $\mathbf{\Sigma}$.
In principle, one could approximate $\Sigma_{i,j}$ using inter-channel spatial correlations averaged over time, which will produce weights analogous to background error covariances used in 3DVAR DA.  
A more fundamental intuition of M-Loss in the context of optimization is provided in supplementary text S2.

The matrix (Figure \ref{fig:1}) reveals strong, mostly positive vertical correlations across all 3D prognostic variables ($T, S, U,$ and $V$). 
Significant $T$–$S$ coupling is also evident in the deeper layers ($120$--$500$ m), depicting the stable covarying T-S structure below the mixed layer and in deeper water masses. 
$SSH$ exhibits distinct positive correlations with temperature across multiple depths and with zonal velocity ($U$) in the upper $100$ m, but a slight anti-correlation with meridional velocity ($V$) over the same depth range. 
Because these correlations are averaged globally and over multiple years, they represent a climatological physical pattern, averaging out localized seasonal or geographic patterns such as specific upwelling/downwelling regimes.

To isolate the impact of the optimization using this $\mathbf\Sigma$, we compare the following two emulator configurations,
\begin{itemize}
    \item Emulator-24h-MSE: Optimized using standard weighted Mean Squared Error, i.e., where $\mathbf\Sigma = I$.
    \item Emulator-24h-MLoss: Trained using M-Loss with $\mathbf\Sigma$ as shown in Figure \ref{fig:1}. 
\end{itemize}

A critical distinction from the original GraphCast configuration is that none of these utilize multi-step autoregressive rollout during the training. 
This may help reduce spectral blurriness and increase physical robustness by allowing more small-scale variability.
In the results, we use Persistence as a baseline, to compare the results against a zero-cost static ocean. 

\section{Results}   

\subsection{Forecast Skill}
\begin{figure}[ht]
    \centering
    \includegraphics[width=\textwidth]{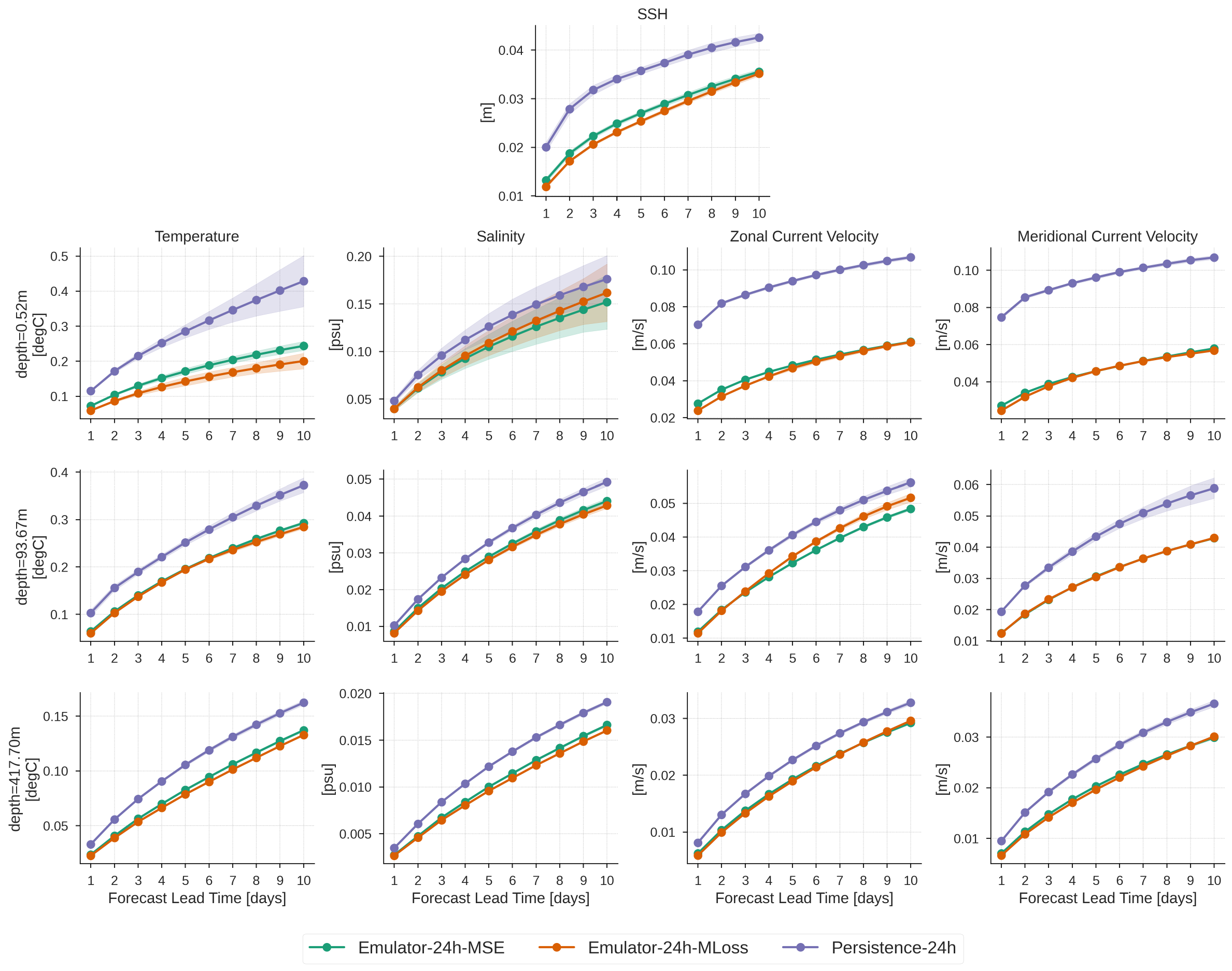}
    \caption{Global RMSE as a function of lead time for SSH (top panel) and 3D prognostic variables: potential temperature, salinity, and zonal/meridional velocities (columns 1–4). 
    The 3D variables are shown at representative depths of approximately $0.5$ m (row 2), $94$ m (row 3), and $418$ m (row 4) corresponding to the surface, the mixed layer, and the deeper water mass. 
    The RMSEs are aggregated over forecasts using $292$ initial conditions in 2022 with a 30 hour cadence.
    Shaded regions denote 95\% confidence intervals estimated using Student's t-test with autocorrelation-based sample size inflation following \citeA{wilks2011statistical} and \citeA{geer2016significance}. A wider CI for SSS reflects its slower temporal evolution, resulting in strongly autocorrelated errors and a reduced effective sample size.}
    \label{fig:2}
\end{figure}

The predictive skill of the two emulators and the Persistence baseline, measured by the global-mean Root Mean Square Error (RMSE), over $10$ days lead time is shown in Figure \ref{fig:2} (mean absolute error (MAE) and anomaly correlation coefficients are shown in Figures S3--S4). 
\noindent Both Emulator-24h-MSE (green) and Emulator-24h-MLoss (orange) significantly outperform the Persistence baseline (purple) across all variables and depths (all depths shown in Figure S2), demonstrating the efficacy of graph-based architecture in emulating global oceans.
Furthermore, the MLoss-based emulator exhibits clear performance gains over the MSE-based emulator in several key fields and depths.
These gains are most pronounced in the surface layer ($\approx 0.5$ m) and the mixed upper ocean ($\approx 94$ m) and diminishes at the deeper oceans ($\approx 418$m), where the MLoss-based emulator converges to the MSE-based emulator with equal or better performance.
The strongest gain is seen for the Sea Surface Temperature (SST), for which the orange curves consistently lies well below the green curve, with the gap widening as the lead time progresses.
At the $10$-day lead time mark, a skill gain equivalent to almost $3$ day is evident for SST using M-Loss compared to MSE.
For SSH, the strongest gain using M-Loss is seen around the $4-5$ day lead time mark, and the gap closes out with increasing lead time.
Both SST and SSH are important surface ocean dynamical variables and are regularly used as boundary conditions in atmosphere-only models.
Improving these quantities could provide better boundary conditions for atmosphere-only models, which may help improve the tightly coupled atmospheric quantities such as 2-meter temperature and humidity.
Improvements in the surface ocean current velocities are also evident up to $5$ day lead time, after which the MSE and M-Loss error curves converge.

Except for sea surface salinity (SSS), which exhibits potential degradation in RMSE skill using the M-Loss (discussed below), an overall improvement in the surface variables suggest that the correlation-aware loss better constrains the surface dynamics by respecting the multivariate dependencies identified in the covariance matrix (Figure \ref{fig:1}).
At deeper levels, the current velocities exhibit a minimal impact of the M-Loss except a degradation in zonal current velocities at lead times larger than $3$ days at around $94$m depth; a full 3D skill comparison plot in Figure S5 suggests a degradation in U-velocity around $100-300$m.
A degradation around this depth is most likely related to the distinct dynamics related to mixing and sensitivity to seasonally-varying surface conditions that may not be well captured by a single aggregated correlation matrix shown in Figure \ref{fig:1}.
We hypothesize that a geographically and/or seasonally varying inter-channel correlation matrix (e.g., separate $\Sigma$ for each geographical location or season) would help resolve these regions better, but exploring this is beyond the scope of this paper. 
An overall alignment of the $U$ and $V$ error curves for M-Loss and MSE losses in the deeper levels suggests that the models reach a similar dynamical steady state, though the M-Loss model's ability to maintain higher surface skill indicates a more physically consistent representation of the wind-driven and geostrophic circulations.
Overall, an improvement in the surface layer quantities and minimal impact in the deeper layers could be related to better constraining of surface quantities due to important mixed layer feedback, whereas deeper level quantities become independent and less impacted by the surrounding layers leading to a convergence between MSE and M-Loss.
\begin{figure}[ht]
    \centering
    \includegraphics[width=\textwidth]{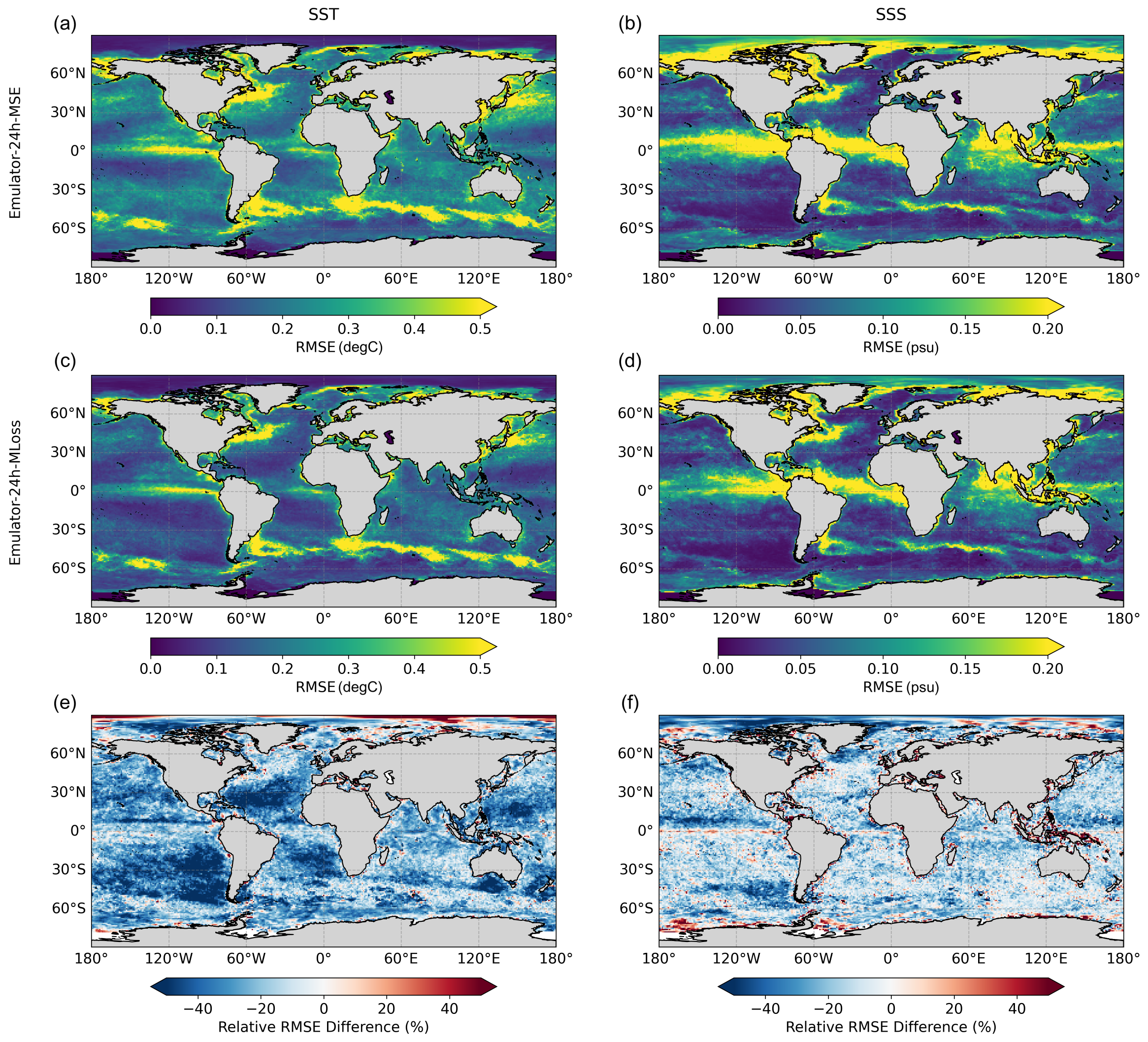}
    \caption{Comparison of 10-day lead time RMSE maps for SST (a,c,e) and SSS (b,d,f) from the two emulators: (a-b) RMSE maps from the MSE-based emulator; (c-d) RMSE maps from the MLoss-based emulator, and (e-f) relative percentage difference in RMSE (M-Loss minus MSE) with MSE as the baseline. Blue/red percentage differences suggests gain/degradation in RMSE skill using M-Loss.}
    \label{fig:3}
\end{figure}

To understand the degradation in RMSE in the SSS using M-Loss compared to the MSE loss, increasing with the lead time, we plot and compare the spatial maps of $10$-day forecast RMSEs for SST and SSS, where SST benefits significantly from the M-Loss at this lead time in contrast to SSS's degradation in the skill (Figure \ref{fig:3}).
It is encouraging to see that M-Loss leads to improvements in a large swath of spatial locations in SST, with more than $50\%$ improvement in RMSEs for locations in the subtropical gyres of Atlantic and Pacific oceans.
We posit that these improvements are due to the slow gyre dynamics and their dependence on the evolution of other important ocean quantities, captured by the correlation matrix in Figure \ref{fig:1}.
Despite the fact that the M-Loss improves the SSS pattern in several regions across the globe, the magnitude is lower, and the improvement in the gyres is not as pronounced as is observed for SST.
Additionally, in the tropics, e.g., the tropical Pacific and maritime continent, a band of red values is visible -- signifying a reduction in the performance using M-Loss.
Similar areas of degraded skill are also visible at high latitudes in Arctic, Antarctic, and near the Gulf Stream extension region. 
A net higher 10-day RMSE for SSS using M-Loss shown in Fig 2 is a result of these localized regions of degraded performance.
This is further confirmed by a comparison of MAE for these emulators (Figure S3), where SSS shows a lower MAE for M-Loss compared to MSE. 
It is worth noting that M-Loss also leads to significant improvements in certain regions in the Arctic  ($30^{\circ}-120^{\circ}$W) in both SST and SSS.
However, a more accurate assessment of the forecast skill for these regions is only possible when sea ice variables are also included in the emulator -- a planned future extension of this work.  

\subsection{Improved Physical Realism}
\label{subsec:improved_physical_realism}

Beyond the traditional numerical skill measured by global mean error metrics and anomaly correlation coefficients, we assess the physical realism of the emulators by evaluating their ability to preserve the fundamental inter-variable spatial correlations. 
In traditional numerical models, these correlations are enforced by the underlying physical and balance equations, e.g., momentum, thermodynamical, hydrostatic balance equations, and the equation of state. 
In data-driven emulators, however, these relationships must be learned from the training data and can be easily disrupted depending on the quality of the training, loss function, architecture, etc.
Such inter-variable spatial correlations, and their alignment with reference truth datasets, have been investigated in the context of atmospheric emulators \cite{schreck2024community}.

\begin{figure}[ht]
    \centering
    \includegraphics[width=\textwidth]{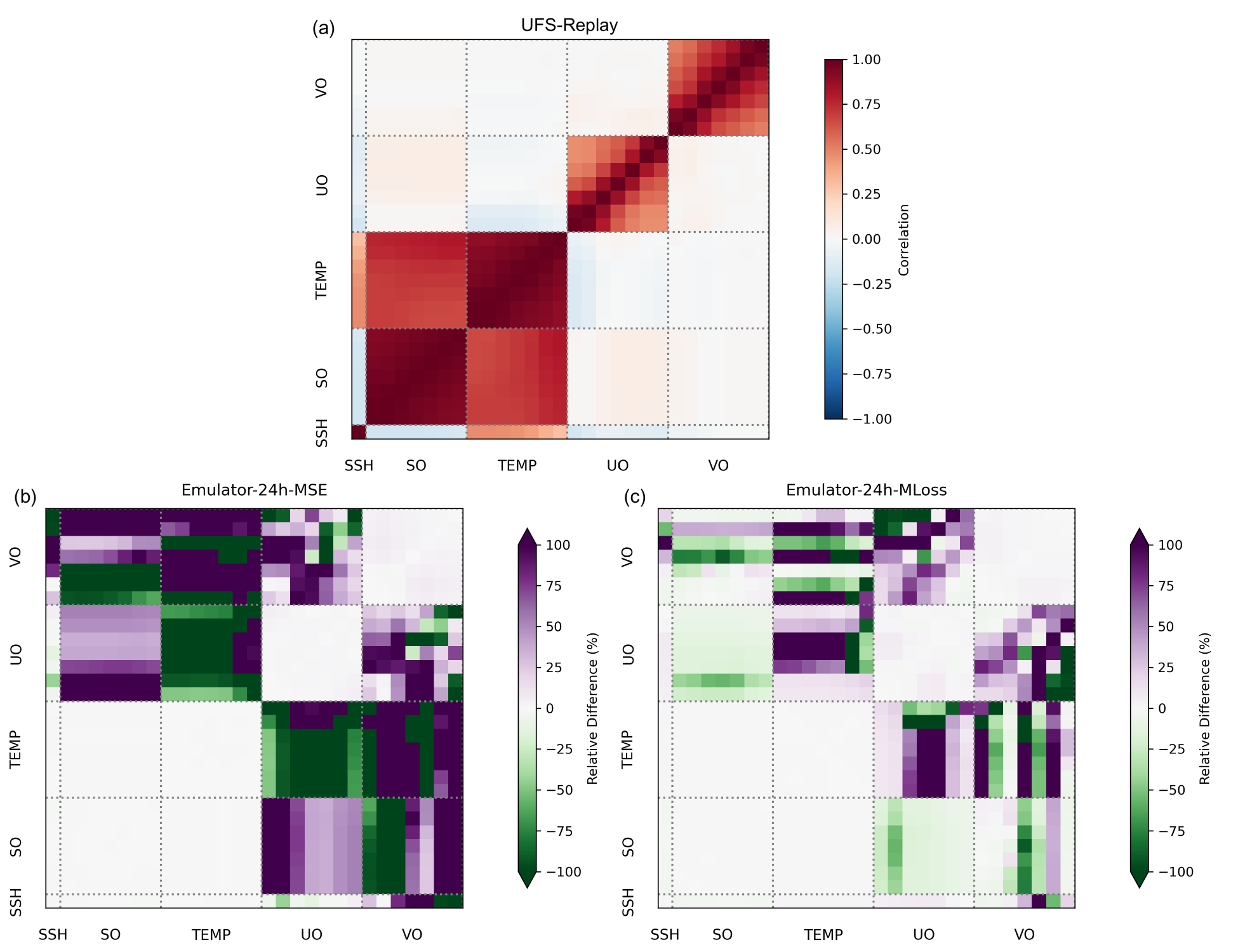}
    \caption{Inter-variable spatial correlation analysis for SSH, temperature, salinity, and zonal and meridional current velocities: (a) the Pearson correlation coefficients for the UFS-Replay reference truth, (b) percentage relative error in the correlation coefficients for forecasts from the MSE-based emulator compared to the UFS-Replay reference, (c) same as (b) but for the MLoss-based emulator.}
    \label{fig:4}
\end{figure}

We quantify inter-variable spatial correlations in forecasted model outputs in three steps: (i) flatten the spatial dimensions (lat/lon) into a single axis, (ii) compute Pearson correlation coefficients along this axis for each temporal snapshot, and (iii) aggregate correlations over the temporal dimension to yield a scalar for each pair of variables. 
For this analysis, spatial correlations for both emulators were computed using 10-day lead time forecasts averaged over all initial conditions in $2022$, with the corresponding UFS-Replay states serving as the reference truth. 
Given the high spatial density of each snapshot, these statistics are robust and do not require large temporal ensembles to reach significance.
Figure \ref{fig:4} illustrates the fidelity of the learned physical relationships by comparing inter-variable spatial correlations in 10-day forecasts from both emulators against the ground truth (UFS-Replay).

Clearly, both Emulator-24h-MSE and Emulator-24h-MLoss successfully resolve strong $T$--$S$ spatial coupling and prominent vertical correlations within individual 3D variables. 
This performance is similar to the behavior of state-of-the-art atmospheric emulators discussed in \citeA{schreck2024community}, with $T$--$S$ coupling analogous the robust temperature-humidity correlations found in the atmosphere. 
Both cases represent fundamental covarying relationships resulting from the emulator’s internalized representation of density computations.
The reproduction of these robust inter-variable spatial correlations suggests that the graph-based encoder-processor-decoder architecture effectively learns these statistical characteristics and maintains them even when rolled out autoregressively. 
While such spatial correlations may degrade at longer lead times alongside the overall forecast quality, the network's current ability to preserve these structures suggests that the latent space effectively encodes the most prominent co-varying quantities.

The most significant divergence between the MSE- and MLoss-based emulators occurs in the spatial correlation blocks between salinity ($S$) and zonal/meridional velocities ($U, V$). 
Despite their modest magnitudes ($0.1–0.2$), these are robust statistics informed by the training data that must be preserved, particularly due to their importance in perturbation studies and data assimilation. 
As illustrated in the relative difference maps (Figure 4), the MSE-based emulator exhibits significant overprediction, i.e., spurious inter-variable correlations, errors in these blocks (magnitudes more than $100\%$), whereas the MLoss-based emulator significantly reduces these errors, shifting toward a slight underprediction with relative error magnitudes generally between $0-20$\%. 
Similarly, the MLoss-based emulator achieves notable error reductions in $T$--$U$ and $T$--$V$ correlations despite their smaller magnitudes. 
These striking differences suggest that the M-Loss objective fundamentally impacts the network's internal learning. 
By explicitly penalizing deviations from the prescribed covariance structure, the model is forced to respect a more physically robust multivariate structure rather than prioritizing high-variance strongly covarying properties to minimize the overall error while remaining approximately physically consistent.

\section{Conclusions and Discussion}
In this work, we established a framework for global 3D ocean emulation using graph neural networks and demonstrated performance gains using a correlation-aware loss optimization.
Our results demonstrate that using a single initial condition with a $24$ hour time step performs well with numerically stable multi-month rollout (Figure S6), and the M-Loss brings further skill gains for SST -- up to $3$ days lead time -- and up to 20\% error reduction for 3D variables at different depths but some degradations between $100–300$m.
We attribute this degradation to complex, inhomogenous thermocline dynamics for which a spatiotemporally averaged $\mathbf{\Sigma}$ is not a correct approximation.
The computational overhead of the M-Loss is minimal as the $\mathbf{\Sigma}$ matrix is computed, factorized, and stored offline before computing the loss (Eq. \ref{eq:maha_loss}) as an $L2$ norm of a scaled and rotated error vector.

Using a multivariate spatial correlation analysis of the forecasted outputs, we demonstrate that explicitly incorporating the multivariate covariance structure during training alters the network's internal learning of the dynamics.
A more consistent representation of the multivariate correlation structure offers substantial promise for DA, as these reduce the potential risk of unphysical balances and initialization shocks. 
By ensuring that inter-variable dependencies are physically consistent, the model allows the impact of localized observations in one field to be felt more coherently across all other prognostic variables during the analysis step. 
For example, in the assimilation of surface pressure observations to constrain and reconstruct global states \cite{compo2011twentieth}, in a system with consistent multivariate correlations, an update in the pressure field informatively and correctly propagates to temperature, velocity, and other components. 
This multivariate covariance fidelity also suggests that emulators from M-Loss objective are better suited for future strongly-coupled DA frameworks. 
In these complex multi-component systems, maintaining a balanced background state between the ocean and the atmosphere is essential, and a correlation-aware emulator can provide the right multivariate structure necessary for producing consistent coupled Earth system reconstructions.

A limitation of this study is the use of a static, globally averaged $\mathbf{\Sigma}$. 
While this captures the correlated dimensions on``average'', it suppresses localized or seasonally-varying properties that do not align with the global average.
Therefore, a future extension will focus on implementing a spatially varying $\mathbf{\Sigma}$ (size: lat $\times$ lon $\times$ targets $\times$ targets) to better accommodate the inhomogeneous nature of the global ocean.
Allowing separate covariance structures over space and seasons would resolve different dynamical regimes and dominant oceanic processes, such as mixed layer deepening/shoaling, deep convective regions in high latitudes, turbulent western boundary current dynamics, and coastal variability, within a manageable computational cost. 
Additionally, evaluating these models within a data assimilation framework with sparse observations, e.g., following \citeA{slivinski2025assimilating}, would reveal how effectively each model uses sparse observations (e.g., SST or SSH) to update the full ocean state vector.

\section*{Open Research Section}
The trained emulator weights and the scripts from this study are available for review at \url{https://figshare.com/s/2f4cd18c38c7d1d52beb}. This will be published publicly upon acceptance for publication. The $1^{\circ}$ UFS-Replay dataset used for training and inference is available publicly and all information about it can be found at \url{https://psl.noaa.gov/data/ufs_replay/}. 

\section*{Conflict of Interest declaration}
The authors declare no conflicts of interest for this manuscript.

\acknowledgments
This research was supported by the NOAA Physical Sciences Laboratory and NOAA cooperative agreement NA22OAR4320151, for the Cooperative Institute for Earth System Research and Data Science (CIESRDS). 
This research used resources of the National Energy Research Scientific Computing Center (NERSC), a Department of Energy User Facility using NERSC award AI4Sci@NERSC-ERCAP0035798.
The authors acknowledge insightful discussions with Noah Brenowitz (Nvidia), Matthew Chantry (ECMWF), Rachel Furner (ECMWF), Lorenzo Zampieri (ECMWF), Ian Grooms (University of Colorado Boulder), and Joshua DaRosa (MITRE) regarding this work. 
The statements, findings, conclusions, and recommendations are those of the authors and do not necessarily reflect the views of NOAA or the U.S. Department of Commerce.

\bibliography{references}

\end{document}